\documentclass{iau} 
\usepackage{graphicx}
\usepackage{xcolor}
\def\rfe{$R_{\rm FeII}$}
\usepackage{amsmath}

\title[RL Quasars and the Main Sequence] 
{Optical spectral properties of radio loud quasars along the main sequence} 

\author[A. del Olmo, P. Marziani, V. Ganci et al.]   
{Ascensi\'on del Olmo$^1$, Paola Marziani$^2$, Valerio Ganci$^{2,3}\footnote{Present address: Institute of Physics, University of Cologne, Germany}$, Mauro D'Onofrio$^3$, Edi Bon$^4$, Natasa Bon$^4$ \and Alenka C. Negrete$^5$}

\affiliation{$^1$Instituto de Astrof\'{\i}sica de Andaluc\'{\i}a, IAA-CSIC, Granada, Spain, email: {\tt chony@iaa.es};\\$^2$INAF,  Astronomical Observatory of Padova, Padova, Italy;  $^3$Dipartimento di Fisica e Astronomia, University of Padova, Padova, Italy; $^4$Astronomical Observatory Belgrade, Belgrade, Serbia; $^5$Instituto de Astronom{\'\i}a - UNAM, CDMX, M\'exico}
	
\pubyear{2019}
\volume{356}  
\jname{Nuclear Activity in Galaxies Across Cosmic Time}
\editors{M. Povi\'c, P. Marziani, J. Masegosa, H. Netzer,\\ S. H. Negu, \& S. B. Tessema, eds.}
\begin{document}

\maketitle

\begin{abstract}
We analyze the optical properties of Radio-Loud quasars along the Main Sequence (MS) of quasars. A sample of 355 quasars selected on the basis of radio detection was obtained by cross-matching the FIRST survey at 20cm and the SDSS DR12 spectroscopic survey. We consider the nature of powerful emission at the high-Fe{\sc ii} end of the MS. At variance with the classical radio-loud sources which are located in the Population B domain of the MS optical plane, we found evidence indicating a thermal origin of the radio emission of the highly accreting quasars of Population A.

\end{abstract}

\firstsection 

\section{Introduction}

The 4D Eigenvector 1 (E1) is a powerful tool to contextualize the diversity of observational properties in type-1 AGN(see e.g., Marziani et al. 2018 for a recent review). The distribution of the quasars in the E1 optical plane, defined by the FWHM of the H$\beta$ broad component vs. the Fe{\sc{ii}} strength (parametrized by the ratio \rfe\,=\,I(Fe{\sc ii}$\lambda$4570)/I(H$\beta$)), outlines the quasar Main Sequence (MS). The shape of the MS (Figure 1, left plot) allow us for the subdivision of quasars in two Populations (A and B) and in bins of FWHM(H$\beta$) and Fe{\sc ii} which define a sequence of spectral types (STs):
\\ 
1) Pop. A with FWHM(H$\beta$)\,$\leq$\,4000\,kms$^{-1}$, and with STs defined by increasing \rfe\ from A1 with \rfe\,$<$\,0.5 to A4 with 1.5\,$\leq$ \rfe\,$\leq$\ 2. STs A3 and A4 encompass the extreme Pop. A of highly accreting quasars radiating near the Eddington limit.
\\
2) Pop. B with FWHM(H$\beta$\,$>$\,4000\,kms$^{-1}$, and STs bins (B1, B1$^+$, B1$^{++}$,...) defined in terms of increasing $\Delta$FWHM(H$\beta$)\,=\,4000\,kms$^{-1}$ (see sketch in Figure 1, left plot).

Eddington ratio ${\rm \lambda_E=L_{Bol}/L_{Edd}}$ and orientation are thought to be the main physical drivers of the MS \cite[(Marziani et al. 2001; Sulentic et al. 2017)]{marzianietal01,sulenticetal17}. Eddington ratio changes along the \rfe\ axis. Pop. B quasars are the ones with high black hole mass ($M_{\mathrm{BH}}$) and low ${\rm \lambda_E}$ and Pop. A are fast-accreting sources with relatively small $M_{\mathrm{BH}}$. Radio-Loud (RL) quasars, defined as relativistic jetted sources, are not distributed uniformly along the MS \cite[(Zamfir et al. 2008)]{zamfiretal08}. They are predominantly found in the Pop. B domain, having \rfe\, $<$\,0.5 and FWHM(H$\beta$)\,$>$\,4000\,kms$^{-1}$. The extreme broad Pop. B$^{++}$ quasar bin (FWHM(H$\beta$)\,$>$\,12000\,kms$^{-1}$) contains about 30\% of the RLs but only $\sim$\,3\% of quasars \cite[(Marziani et al. 2013)]{marzianietal13}.
\vspace{-0.5cm}

\section{Sample}
We selected quasars from the SDSS-BOSS DR12 Quasar Catalog \cite[(P{\^a}ris et al. 2017)]{parisetal17} with m$_i\leq19.5$ and redshift\,$\leq$1 and cross-matched with the VLA FIRST radio survey at 20cm \cite[(Becker et al. 1995)]{beckeretal95}. A detailed description of the sample and the performed analysis can be found in \cite[Ganci et al. (2019)]{gancietal19}. The classification of the quasars was carried out according to three criteria:
\begin{itemize}
\item[$-$] Radio power, based on the ${\rm R_\mathrm{K}=f_{rad}/f_{opt}}$ parameter (Kellerman et al. 1989), defined on the basis of the 1.4\ GHz and $g$\ magnitude estimates. Sources with 
${\rm log R_K}$\,$<$\,1.0 are classified as Radio-Detected (RD); those with 1.0\,${\rm \leq log R_\mathrm{K}<}$\,1.8 as radio-intermediate (RI) and for ${\rm log R_\mathrm{K}\geq}$\,1.8 as RL.
\item[$-$]Radio morphology as Core Dominated (CD) and FRII sources, selected using the statistical procedure defined by \cite[de Vries et al. (2006)]{devriesetal06}.
\item[$-$] Optical ST classification in terms of the MS.
\end{itemize}
\vspace{0.05cm}

The final sample consist of 355 objects, 289 CD and 66 FRII. There are a total of 38 RD, 139 RI and 178 RL objects.
\vspace{-0.4cm}

\section{Results}
We have analyzed the source distribution along the MS. Table 1 shows a summary of the number of sources for both Pop. A and B and for the particular case of the extreme accreting Pop. A quasars (xAs, STs A3-A4), according to its radio loudness and radio morphology. From this analysis we found: 
\begin{itemize}
\item[\textbullet]The most populated STs are the Pop. B bins (72\% of the sources), especially B1 (36\%) and B1$^{+}$ (22\%) at variance with optically selected samples \cite[(Marziani et al. 2013)]{marzianietal13} where B1 and A2 are the most populated STs. 
\item[\textbullet] RD sources are only present in CD radio morphology (21 Pop. A, and 17 Pop. B). 
\item[\textbullet] RIs sources have also only CD morphology, but are more numerous in Pop. B (70\%). 
\item[\textbullet] RLs are significantly more numerous in Pop. B for the CD morphology and FRII sources appear only in RL class, and are almost exclusively Pop. B (95\%).
\end{itemize}

The most powerful radio quasars are located in Pop. B both for FRII and CD radio morphologies and they are most likely jetted RL quasars, meanwhile Pop. A shows almost exclusively CD morphology, with about 28\% of the CD objects classified as Pop. A but with significant lower radio power that CD Pop. B sources. 

\begin{table}[t!]
	\begin{minipage}[t]{0.6\linewidth}
     \centering
      \begin{tabular}{lccccccc}\hline 
      {} & \multicolumn{3}{c}{CD} & & \multicolumn{3}{c}{FRII}\\
      \hline
      {} & RD & RI & RL && RD & RI & RL \\ 
      Pop. B & 17 & 96 & 81 && - & - & 63\\ 
      Pop. A & 21 & 43 & 31 && - & - & 3 \\ 
      xA (STs A3-A4) & 5 & 17 & 5 && - & - & - \\
      \hline
      \end{tabular} 
    \end{minipage}
    \begin{minipage}[th!]{0.4\linewidth}
	 \centering
	 \vspace{-0.5cm}
	 \hspace{10cm} 
     \caption{Number of sources by population and radio-loudness.} 
    \end{minipage}
\end{table}
	
The spectral analysis of the SDSS-BOSS spectra of the sample confirms the systematic differences between Pop A and Pop. B, and the trends associated with the quasar MS, similar for all radio classes.

\begin{itemize}
\item[\textbullet] The centroid at half maximum of H$\beta$ passes from redshifted in Pop B, due to the presence of the very broad component  characteristic of this population of quasars, to clearly blueshifted in xAs. 
\item[\textbullet] Equivalent width of H$\beta$ remains roughly constant in Pop. B quasars, but dramatically decreases in Pop. A. 
\item[\textbullet] From the B1$^{++}$ to the A4 STs, ${\rm \lambda_E}$ increases and $M_\mathrm{BH}$ decreases, independently of the radio classes.
\end{itemize}

\vspace{0.2cm}

A very interesting result from our analysis is that about 80\% (22/27) of the xA CD have a radio emission in the loud range (RI and RL). Are these RL and RI xAs truly jetted sources?, or is its radio power dominated by thermal emission? We have investigated the possible origin of their radio emission by analyzing its FIR luminosity and through the FIR-radio star formation rate (SFR) relation (Bonzini et al. 2015), since RL jetted sources are expected to have FIR luminosity significantly lower than the one expected for their radio power. 
We show in Figure 1 (right) the FIR and radio power SFR for  
the xAs sources with FIR data available from both this sample and the xAs in Bonzini et al. (2015) sample (see Ganci et al. 2019), together with the location of the star forming galaxies (gray shaded area) and radio-quiet (RQ) quasars (pink area). 
xAs sources have both high radio power and SFR. They are placed in the RQ and star-forming galaxies regions, suggesting that their radio power is due to the thermal emission. And none of the other sources, included in Table 1, that fall in the RL region (outside and to the rigth of the shaded areas) is an xA source. 

\begin{figure}
	\begin{minipage}{0.5\linewidth}
		\centering
		\includegraphics[width=5.5cm]{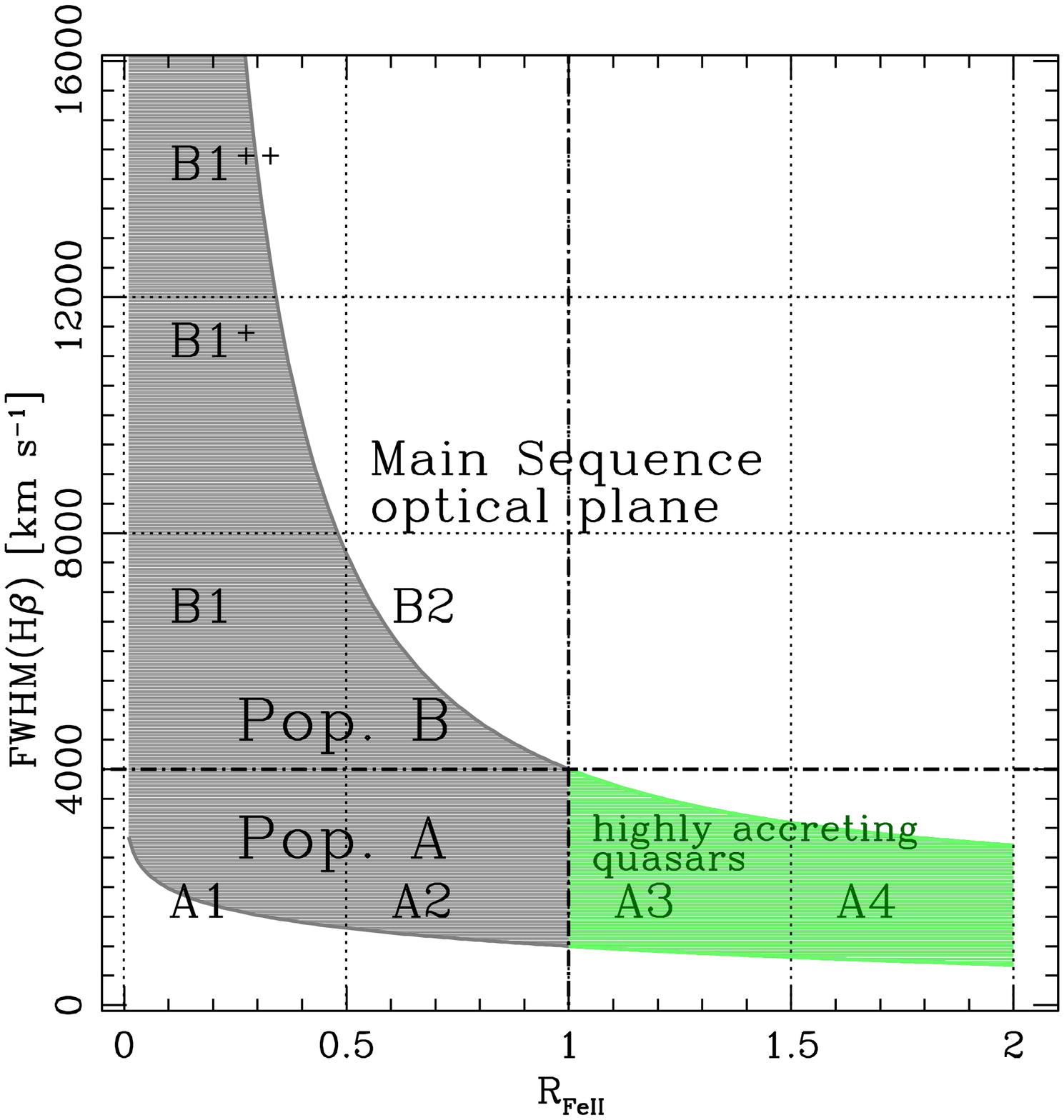}
	\end{minipage}
	\begin{minipage}[h!]{0.5\linewidth}
		\centering
		\vspace{-0cm} 
		\includegraphics[width=5.5cm]{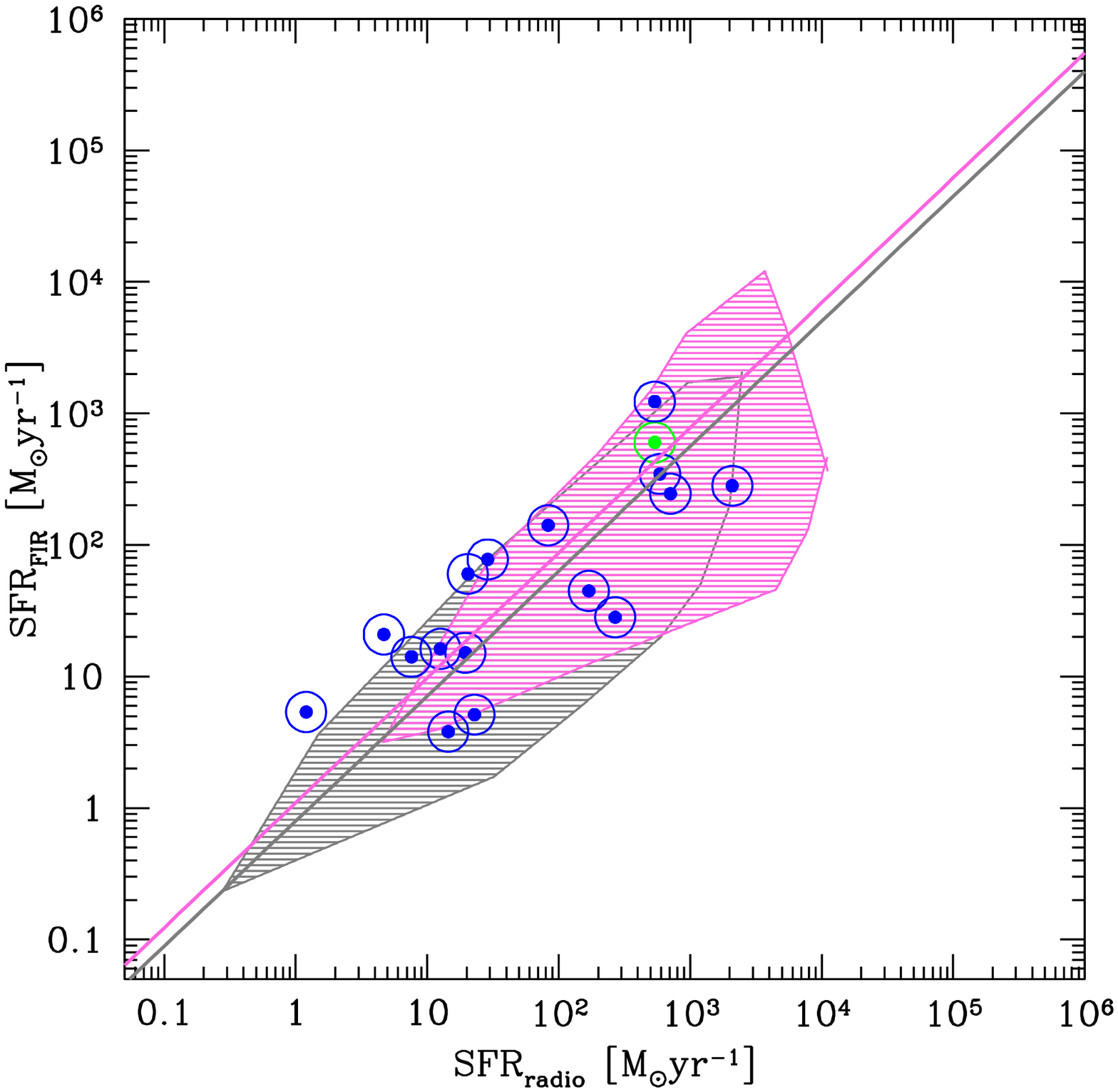}
	\end{minipage}
	{\small
	\caption{Left: sketch of the optical plane of the MS of quasars and the STs. Green area traces the location of the extreme Pop. A quasars (xAs); Right: FIR vs. radio SFR for the xAs with FIR data. The shaded areas trace the loci of star-forming galaxies (grey) and RQ quasars (pink). }  }
\end{figure}

It is therefore reasonable to suggest that most of the xAs that are RI and RL are non-jetted sources and they might be truly “thermal sources”. Our study (Ganci et al. 2019) supports the suggestion that RI and RL A2, A3, A4 sources could be thermal in origin.

\vspace{0.2cm}
{\bf Acknowledgements.} {AdO acknowledges financial support from the Spanish grants MEC AYA2016-76682-C3-1-P and the State Agency for Research of the Spanish MCIU through the ``Center of Excellence Severo Ochoa" award for the IAA (SEV-2017-0709).}

\end{document}